# Comparative Study on DFD to UML Diagrams Transformations

Atif A. A. Jilani
National University of Computer and Emerging Sciences, Islamabad
atif.jilani@nu.edu.pk

Muhammad Usman
Mohammad Ali Jinnah University Islamabad
m_usman99@hotmail.com

Aamer Nadeem
Mohammad Ali Jinnah University Islamabad
anadeem@jinnah.edu.pk

Zafar I. Malik
Academy of Education & Planning, Ministry of Education, Pakistan
zafarimalik@hotmail.com

Zahid Halim
National University of Computer and Emerging Sciences, Islamabad
zahid.halim@nu.edu.pk

Abstract—Most of legacy systems use nowadays were modeled and documented using structured approach. Expansion of these systems in terms of functionality and maintainability requires shift towards object-oriented documentation and design, which has been widely accepted by the industry. In this paper, we present a survey of the existing Data Flow Diagram (DFD) to Unified Modeling language (UML) transformation techniques. We analyze transformation techniques using a set of parameters, identified in the survey. Based on identified parameters, we present an analysis matrix, which describes the strengths and weaknesses of transformation techniques. It is observed that most of the transformation approaches are rule based, which are incomplete and defined at abstract level that does not cover in depth transformation and automation issues. Transformation approaches are data centric, which focuses on data-store for class diagram generation. Very few of the transformation techniques have been applied on case study as a proof of concept, which are not comprehensive and majority of them are partially automated.

## I. INTRODUCTION

Most of the legacy systems in use nowadays were modeled and documented using structured approach [1]. These systems were developed in languages that have become out-dated now. With the passage of time, systems demand numerous modifications, expansion in terms of functionality and incorporation of latest high-speed hardware. Still, legacy systems are reliable enough and considered irreplaceable by the user. However, it is possible to modify system code but modifications in the code add inconsistencies between code and design and system design becomes no longer usable for future maintenance. Besides, modifying such a system is also very costly, the only viable solution for up-gradation and maintenance is to preserve system design and incorporate it with latest software development strategies as described by Newcombe and Doblar [2].

If running system code is available, it is possible to generate design from code. However, if the code is modified numerous times, the generated design and original design may become inconsistent. Design recovery from such code is ambiguous and no more useful for future up-gradation and maintenance. Like Dietrich et al. [3], we also consider legacy systems irreplaceable and trusted by the users. We, too, emphasize on saving legacy system by providing and using an object-oriented interface.

A major design artifact in structured approach is the Data Flow Diagram (DFD). Other artifacts like structure chart, state machine, and ER diagram are also there, but DFD has certain advantages over them. DFD is the primary artifact and is required be created for every systems in structured approach. DFD has hierarchal structure, which provides different abstraction level, useful in system designing.

Besides, DFD is such a fundamental artifact that clearly depicts the structure of a system. Other artifacts use the information provided by the DFD to represent dynamic aspect of the system [3].

Structured design techniques have been replaced by object-oriented analysis and design approach, which has gained popularity now and majority of the software modeling and development techniques are adopting this paradigm [4]. With the passage of time, the level of abstraction in system development has raised. Object Management Group (OMG) has been recently promoting a new vision for software development, i.e., Model Driven Architecture (MDA) [5]. In MDA, main emphasis is on modeling design separately from the implementation (platform). MDA encourages the use of Platform Independent Model (PIM) and Platform Specific Model (PSM) to represent platform independent and platform specific details. Soul of MDA is transformation between models. In MDA, code may be generated either from PSM or from the PIM. Model transformations can be between PIM and PSM, PIM and code or between PSM and code. For modeling object-oriented systems and model creation, Unified Modeling Language (UML) [6] has now become the de-facto industry standard [7-8]. UML is a collection of diagrams used to model different aspects of object oriented software. UML Class diagram is one of a major artifact in object-oriented design used to represent the system's static structure. Other UML diagrams, like sequence diagram, state machine, and activity diagram etc, are used to model the system's dynamic behavior.

In this paper, we present a survey of transformation techniques that are used to generate legacy system design in UML. We include DFD to UML diagrams transformation in our survey. We analyze different existing transformation techniques using set of Analysis parameters identified in the survey. Based on the parameters analysis matrix is created, which highlights the weaknesses and strengths of different techniques. Motivation behind DFD-UML models transformations is that designers/analysts can use surveyed transformations from DFD to class diagram, with the existing MDA transformation [9] either as PIM to PSM or as PIM to code.

## II. Scope of the Survey

Modernization of legacy systems cost effectively has become the primary focus of software designers and researchers. In literature, both structured code to object-oriented design and structured design to object-oriented design transformations exists. Code to design techniques cause maintenance issue because code is written in programming languages, which have become out-dated now. Besides, code might have undergone numerous modifications, which make code and design inconsistent. Design recovery from that code is useless and no more usable for future maintenance. Contrary to this, in structured design to object-oriented design transformations, new design provides basis for development of new system and becomes future reference for maintenance. We limit our scope to legacy system design to UML design. In particular, we focus only on data flow diagram to UML design transformation.

In literature, both structured code to object-oriented design and structured design to non-UML object-oriented design techniques exists. In this section, we discuss both the views.

### A. Structured Code to Object-oriented Design Techniques

In structured code to object-oriented design techniques Liu and Wilde [9], propose methodologies for identifying object from non-object-oriented languages. They propose type base and global base object finder methodologies. Jacobson and Lindstrom [10] describe reverse engineering strategies and discuss object-oriented model to incorporate changes. Livadas and Johnson [11] propose an approach that maintains existing relationship in the maintained code. Similarly, another approach by Gall and Klösch [12] defines two types of data entities: data store entities and non-data store entities. They describe relationship between the two types for expressing entities as objects.

Newcombe and Kotik [13] present a tool for abstract object-oriented model generation. Subramanian and Bwirne [14] generate objects from FORTRAN code. They discuss constraints like private, virtual, and pure virtual. Cimitile et al [15] and De Lucia et al [16] present approaches that revolve around data stores. Authors propose approaches that consider functions and subroutines, interacting with tables, data-store and use them as objects methods. Similarly, De Lucia et al [17] propose an approach that recover class diagram from data intensive legacy system code.

It is apparent that many of the discussed techniques are effective only for data centric systems. For our approach, we are firmly interested in structured design instead of code. In design, we have observed that in literature, both structured design to non-UML design and structured design to UML design transformations exist. We briefly explain both the views.

### B. Structured Design (DFD) to a Non UML Object-oriented Design

In this section, we will discuss those techniques that transform structured design to a non-UML object-oriented design. Alabiso [18] use FDC (Functional Design Chart) to express functional behaviors and OSC (Object Structure Chart) to express breakdown of data structures of object. His transformation approach is not automatable and does not provide detail transformation rules. George and Carter [19] propose mapping strategy that uses Entity Relationship Diagram (ERD), Functional Data Flow Diagram (FDFD) and Data Dictionary as a source model and generates Object Structure and Mapping Diagram (OSMD). Their approach too is not automatable.

## III. Criteria for Evaluation

In this section, we describe and discuss set of analysis parameters. Analysis parameters provide criterion to evaluate different techniques.

For parameters selection, we comprehensively consider transformation techniques description, their limitations and comparison discussed by the authors while describing their respective technique. On reviewing different techniques, certain parameters are identified. Detailed description of parameters and their values is given below.

### A. Automatable

This parameter describes whether a transformation technique is automated or can only be applied manually. The parameter value depicts practical importance of the technique and used in determining the efficiency and applicability of the approach. The values assigned to the parameter are 'Yes', 'No' and 'Partial'. Table 1 shows value selection criteria for automatable.

Table 1 Evaluation criteria for Automation.

| Value | Criteria |
|---|---|
| Yes | Technique automatically transform source model into target model. |
| No | Technique does not automatically transform source model into target model. |
| Partial | Author explicitly mentioned or after analysis of case study and transformation rules, we have found that some manual support is needed for automation. |

### B. Tool Support

Tool support parameter describes whether a tool or an automatable environment is available for the technique or not. Table 2 shows value selection criteria for tool support.

Table 2 Evaluation criteria for Tool Support.

| Value | Criteria |
|---|---|
| Yes | Author explicitly mentioned tool support. |
| No | No information related to tool is provided. |

### C. *Additional Artifact Used*

This parameter describes the additional artifacts used by the transformation technique. Values for this parameter include Data Dictionary (DD), Entity Relationship Diagram (ER) etc.

### D. *Output Artifact*

This parameter describes UML artifacts that are generated because of transformation technique. Values for this parameter could be Class, Use-Case, Sequence, state chart diagram etc.

### E. *Case Study*

This parameter defines whether a transformation technique applied on a case study or not. This parameter is important because a technique needs to be applied on case study to check it applicability. Table 3 shows value selection criteria for case study.

TABLE 3 EVALUATION CRITERIA FOR CASE STUDY.

| Value | Criteria |
|---|---|
| Yes | Case Study provided or discussed. |
| No | No Case study discussed. |

### F. *Transformation level*

This parameter defines type of transformation given technique follows. Value for this parameter includes rule-based or metamodel-based transformation. Table 4 shows value selection criteria for transformation level.

TABLE 4 EVALUATION CRITERIA FOR TRANSFORMATION LEVEL.

| Value | Criteria |
|---|---|
| Rule-based | Transformation rules are provided for models instances. |
| Metamodel-based | Transformation mappings and rules are provided at metamodel level. |

### G. *UML Conformance*

This Parameter describes the conformance of the generated model. It shows model produce by transformation technique follows UML syntax and semantics or not. Table 5 shows value selection criteria for UML conformance.

TABLE 5 EVALUATION CRITERIA FOR UML CONFORMANCE.

| Value | Criteria |
|---|---|
| Yes | Generated model follows UML syntax and semantics, describe in UML Superstructure. |
| No | Generated model do not follow UML syntax and semantics. |

### **H.** *Scalability*

This parameter defines whether a transformation technique has potential, to be applied on larger size case study or not. Case study description provides the basis for this parameter. Table 6 shows value selection criteria for scalability.

TABLE 6 EVALUATION CRITERIA FOR SCALABILITY.

| Value | Criteria |
|---|---|
| Yes | Size of case study is appropriate and technique fully transformed source model to target model in feasible time. Technique description helps in determine in efficiencies. |
| No | Size of case study is small and approach does not fully transformed model in feasible time. |

### I. *Direction*

This parameter defines transformation direction. Values include unidirectional and bidirectional. Table 7 shows value selection criteria for direction.

TABLE 7 EVALUATION CRITERIA FOR DIRECTION.

| Value | Criteria |
|---|---|
| Bidirectional | Technique transforms structured model to object-oriented model and can retransform object-oriented model to structured model. |
| Unidirectional | Technique transforms structured model to object-oriented model only. |

### J. *Input Model Coverage*

Input model coverage parameter describes the coverage of different constructs of source model. Since, our primary focus is on DFD, we check that different constructs of DFD like external entities, data-stores, processes and data-flows between them are properly transformed into target model or not. Table 8 shows value selection criteria for input model coverage.

TABLE 8 EVALUATION CRITERIA FOR INPUT MODEL COVERAGE.

| Value | Criteria |
|---|---|
| Low | Basic constructs like external entities, data-stores, processes are only transformed. |
| Medium | Few data-flows between components along with basic constructs are transformed |
| High | All data-flows between components along with all the constructs are transformed |

### K. *Target Model Coverage:*

Target model coverage parameter describes the coverage of different constructs of the target model. Since, we focus on UML models this parameter describe construct of different UML models. For class diagram, construct of class diagram like owned attribute, owned operation, association, dependency, inheritance relationship are considered. Class diagram metamodel describe in UML superstructure, helps in identifying different constructs. Table 9 shows value selection criteria for output model coverage.

TABLE 9 EVALUATION CRITERIA FOR TARGET MODEL COVERAGE.

| Value | Criteria |
|---|---|
| Low | Basic constructs, like owned attributes and owned operation are catered. |
| Medium | Association between components along with basic constructs is catered. |
| High | All the constructs including inheritance are catered. |

### L. *DFD Level Used:*

This parameter defines the level of DFD used in transformation technique. Values for this parameter include context level DFD and Extended level DFD. Table 10 shows value selection criteria for DFD level used.

TABLE 10 EVALUATION CRITERIA FOR DFD LEVEL USED.

| Value | Criteria |
|---|---|
| Context Level | DFD in transformation has hierarchy which is not refined |
| Extended Level | DFD in transformation is refined and no hierarchy exists. |

Based on the above-mentioned criteria, we evaluate the techniques explained below.

## IV. STRUCTURED DESIGN TO UML DESIGN TRANSFORMATION TECHNIQUES

In this section, we will analyze structured design techniques that generate UML Object-oriented design based on evaluation criteria discussed earlier.

### A. *Documentation Maintenance: DFD by Means of UML. [20]*

In documentation maintenance DFD by means of UML, authors generate several UML diagrams from DFD. UML diagrams generated from DFD are Use-Case Diagram, Class Diagram and Interaction Diagram. Context diagram of DFD transform into Use-Case diagram by mapping data stores and sinks into actors, processes into use cases and data flows into relationships between processes. For class Diagram generation, data-stores are considered as classes, processes attached to the data stores are considered as functions and all the data elements of data store expressed as attributes of classes. For interaction diagrams, processes are used as relationship between generated classes.

### B. *Meta Model Approach for Mediation [21]*

In metamodel approach for mediation, authors propose formal DFD metamodel. DFD metamodel describes DFD semantics formally. DFD instance can be created by translating semantics describe in metamodel. Authors also discuss generation of UML models using DFD metamodel. UML models that transform using DFD metamodel are Use-Case diagram, Class diagram and Sequence diagram. Incomplete and informal mapping rules are also proposed, which are used for DFD to UML models transformation. Shiroiwa proposed metamodel is a mediator between DFD and UML. metamodel also preserves DFD hierarchy structure.

The technique proposed by author is generalized as it is based on metamodel. Case study and tool support description is not provided. Generated UML models conformance UML semantics. Context level DFD is used for transformation. Both the source and target models have medium coverage because transformation technique does not provide rules for every construct of source and target models. Only basic constructs of source model are transformed into the basic constructs of target model.

### C. *Framework for transforming Artifacts of DFD to UML. [22]*

In framework for transforming artifacts of DFD to UML, Tran et al, propose a framework that works on DFD at three levels of abstraction. At DFD abstraction level 1, framework translates DFD into UML Use-Case diagram. At DFD abstraction level 2, framework translates DFD into UML interaction diagram. Instead of DFD, Entity Relationship Diagram (ERD) is used as an additional artifact for generation of UML class diagram structure. At abstraction level 1, processes in DFD are mapped to Use-Cases, external entities are mapped as actors, and data stores are mapped to classes. Data flows mapping is remained unresolved. At abstraction level-2, data flows variations includes external entity to process, data store to process, process to data store and process to external entity are transformed into interaction diagram. Data flow variations process to process are mapped on state transition diagram. At abstraction level 3, ERD components including entity, association entity, attributes, and relationships map to class association, attributes and operations. It is also mentioned by the authors that framework does not address in-depth issues of transformation as complete transformation rules are not provided.

The proposed framework generates valid UML models. It is clearly describe by the author that technique is not scalable. No tool and automation issues for framework are discussed. Technique describes informal transformation rules and is unidirectional. Technique is not practical, because it is partially automatable and no case study is discussed. Every construct of source model is catered in transformation rules, which makes source model coverage high. Transformation rules for basic construct of target model are provided, which makes target model coverage Medium.

### D. *Functional and Object-oriented Views in Embedded Software Modeling. [23]*

In Function and Object-oriented Views in Embedded Software Modeling, Fernandes and Lilius describe DFD and UML diagram transformation. They propose that DFD is used in an integrated way to refine UML models including Use-Case and Class Diagram. They also use DFD to detail the behavior of a system component. Authors describe that since, DFD are more expressive to represent user requirement, as compare to use case diagram it should be used to represents user requirements. DFD transformation to UML diagrams at context level is only applicable. Similarly, for sequence, collaboration and class diagram DFD can also be used.

Technique proposed by authors is partially automatable. Technique generates valid UML models. Technique uses informal rules. Rules for very few construct of source and target model are provided, which make both source and target model coverage low.

### E. *Tool Support for DFD-UML Model-based Transformation [24]*

In the paper tool support for DFD-UML, model-based transformation, authors propose an approach that combines both functional and object-oriented models for modeling embedded system. They also implemented a tool for transformation between different views. Truscan et al, propose Software Modeling Workbench (SMW) that gathers requirements, create use-case diagram and transforms it into non-UML, so call Initial Object Diagram (IOD). SMW also transforms DFD’s into Class Diagrams. Through transformation scripts, basic rules are implemented to perform transformation. Rules are specific only for IPv6 case study.

In this paper, SMW tool discussed by authors, run script to generate class diagram. Proposed technique is automatable and mapped only on specific case study. We consider proposed approach impractical for others systems as of specific rules for IPv6 case study.

### F. *Systematic Transformation of Functional analysis into Object-Oriented Design and Implementation [25]*

In systematic transformation of functional analysis into Object-Oriented Design and implementation, the authors proposed an enhanced data flow diagram called DF net, which is used to specify use cases from requirements. The proposed DF net is also used in transformation to generate object-oriented design. According to authors, the transformation between DF net is carried in different steps. During first step, processes in DF net dealing with data stores, data buffers and external entities are grouped together. Similarly, processes that share the same data such that one process output is the input of other process are grouped separately. Next step is to generate classes from the separated group. Similarly, use-cases are also identified by using DF net. The whole process is automatable and tool support is available for transformation. The proposed approach according to authors is scalable and practicable.

### *G. A Framework for Transformation Structured Analysis and Design Artifact to UML [26]*

In the paper, framework for transformation structured analysis and design artifact to UML, Fries converts DFD and Entity Relationship Diagram (ERD) into UML models. UML models include use-case diagram, sequence diagram, state machine diagram and class diagram. DFD Process in level-1 DFD is mapped on use-case, external entity creates an actor in use-case and data flow creates association line in use case diagram. Similarly, for sequence diagram different data flows are mapped on sequence diagram. State machines are generated by transforming data flows between processes as event parameters. Number of processes defines number of states. ERD is used for creation of class diagram.

The proposed framework is partially automatable. Author partially defines transformation rules that generate valid UML models. Transformation is not practical, as of partial automation, though it is applied on a case study but incomplete rules are provided. Transformation is not scalable because transformation rules are defined at very abstract level. Transformation technique is partially automatable as rules are incomplete. Coverage for the source model is high as it caters every construct of DFD but coverage for target model is medium as basic constructs are only catered for target model.

Table 11 shows the comparison of all the techniques based on the evaluation criteria explained in section 3.

## V. CONCLUSION OF SURVEY

From the comparison of the existing DFD-to-UML design transformation techniques (Table 4.1), we conclude that most of the presented approaches are rule-based and are incomplete; do not supports model validity and model generalization. Techniques in majority of approaches define abstract rules for transformations, which do not cover in depth transformation and automation issues. Few of the techniques have been applied on case study as a proof of concept. Although few authors discuss process-to-process component data flow but none of them proposes appropriate solution for their transformation.

It is observed that rules are defined, but at abstract level, which does not cover in depth transformation and automation issues. Majority of transformation techniques are unidirectional, lacks automation. Few of the transformation techniques have tool support. Some of the techniques have been applied on case studies, which are not comprehensive and majority of them are partially automated. Very few techniques provide high source and target model coverage. We have found that only one metamodel-based technique exists, which too discussed transformation at very abstract level.

It is also observed from the literature review that different analysts/designers have their own interpretation of different DFD graphical symbols as DFD has informal syntax. Transformation approaches are data centric, which focuses on data-store for class diagram generation. Although authors discuss process-to-process component data flow but none of them proposes complete solution for their transformation. ER-diagram is also used as an additional artifact to generate class diagram in some of the techniques.

The comparison in Table 4.1 shows that a solution for DFD to UML class diagram transformation is needed. A solution that will cover in-depth transformation issues by providing detailed transformation rules. Besides, transformation should provide solution for the sequence that transformation follows and cater data flow transformation between processes. Transformation should follow MDA transformation strategy because it is the latest initiative of OMG for model transformations. By following MDA transformation strategy one can use the technique with the existing MDA and Model Driven Engineering (MDE) transformation approaches. Transformation should generate UML class diagram, which is the major artifact in object-oriented design to represent system static structure. Proposed transformation should provide reusable modern object-oriented design that will be helpful for future maintenance. Transformation strategy should be based on formal DFD metamodel, so that DFD design ambiguity and inconsistency is not reflected in generated class diagram.

TABLE 11 ANALYSIS OF THE EXISTING DFD-TO-UML MODEL TRANSFORMATION TECHNIQUES

| **Transformation Technique Name** | **Analysis Parameter** | | | | | | | | | | | |
|---|---|---|---|---|---|---|---|---|---|---|---|---|
| | **Automatable** | **Tool Support** | **Additional Artifact Used** | **Output Artifact** | **Case Study** | **Transformation Level** | **UML Conformance** | **Scalability** | **Direction** | **Input Model Coverage** | **Target Model Coverage** | **DFD Level Used** |
| Documentation maintenance: DFD by means of UML | Partial | No | NIL | Use-Case, Class & Interaction Diagram | No | Rule Based | Yes | No | Unidirectional | Medium | Medium | Context |
| Meta-Model Approach for Mediation | Yes | No | NIL | Class Diagram. | No | Meta-Model Based | Yes | No | Unidirectional | Medium | Medium | Context |
| Framework for transforming Artifacts of DFD to UML. | Partial | No | ER Diagram | Use-Case, Class, Sequence & State Chart Diagram | Yes | Rule Based | Yes | No | Unidirectional | Medium | High | Context |
| Functional and Object-oriented Views in Embedded Software Modeling. | Partial | No | NIL | Class, Use-Case & Interaction Diagram | No | Rule Based | No | No | Unidirectional | Low | Low | Context |
| Tool Support for DFD-UML Model-based Transformation | Yes | Yes | NIL | Class Diagram | Yes | Rule (script) Based | No | No | Bidirectional | Medium | High | Extended |
| Systematic Transformation of Functional analysis into Object-Oriented Design and Implementation | Yes | Yes | DF Net | Class Diagram | Yes | Rule Based | Yes | Yes | Unidirectional | Medium | Medium | Extended |
| A Framework for Transformation Structures Analysis and Design Artifact to UML | Partial | No | ER Diagram | Class, Use-Case, Sequence & State Chart Diagram | Yes | Rule Based | Yes | No | Unidirectional | Medium | High | Context |